# Danish DPA Banned the Use of Google Chromebooks and Google Workspace in Schools in Helsingør Municipality

**Marcelo Corrales Compagnucci**[∗]

## I. Introduction

On July 14th, 2022,[1] the Danish Data Protection Authority (*Datatilsynet* – Danish DPA) issued a reprimand against Helsingør Municipality. It imposed a general ban on using Google Chromebooks and Google Workspace for education in primary schools in the Municipality. The Danish DPA banned such processing and suspended any related data transfers to the United States (U.S.) until it is brought in line with the General Data Protection Regulation (GDPR).[2] The suspension took effect immediately, and the Municipality had until August 3rd, 2022, to withdraw and terminate the processing, as well as delete data already transferred.[3]

Finally, in a new decision on August 18th, 2022, the Danish DPA has ratified the ban to the use of Google Chromebooks and Workspace. In the eyes of the Danish DPA, the Municipality failed inter alia to document that they have assessed and reduced the relevant risks to the rights and freedoms of the pupils.[4]

---

[∗] Marcelo Corrales Compagnucci is Assoc. Professor at the Center for Advanced Studies in Biomedical Innovation Law (CeBIL), University of Copenhagen (UCPH). Email: marcelo.corrales13@gmail.com
This is a draft version. The final version was submitted to the European Data Protection Law Review (EDPL) (currently under peer review). Special thanks to Rie Aleksandra Walle and Mark Fenwick for their valuable comments and feedback to this article. However, the usual disclaimer applies. Any errors belong to the author.

[1] Based on a decision from September 10th, 2021.
[2] Regulation (EU) 2016/679 of the EP and of the Council of 27 April 2016 on the protection of natural persons with regard to the processing of personal data and on the free movement of such data, and repealing Directive 95/46/EC, OJ 2016 L 119, 1 (General Data Protection Regulation, GDPR).
[3] European Data Protection Board (EDPB), The Danish DPA imposes a ban on the use of Google Workspace in Elsinore municipality, available at: https://edpb.europa.eu/news/national-news/2022/danish-dpa-imposes-ban-use-google-workspace-elsinore-municipality_en (accessed 9 August 2022).
[4] Datatilsynet, 'Datatilsynet Fastholder Forbud I Chromebook-sag' (in English: *The Danish Data Protection Authority Maintains the Ban in the Chromebook case*) (18 August 2022), available at: https://www.datatilsynet.dk/presse-og-nyheder/nyhedsarkiv/2022/aug/ny-afgoerelse-datatilsynet-fastholder-forbud-i-chromebook-sag (accessed 20 August 2022).

This article is structured as follows: section II provides the background concerning the unfolding events after the *Schrems II* ruling. Section III discusses the origins and facts of the Danish DPA case. Section IV examines the reasoning and critical findings of the Danish DPA decision. Finally, section V concludes with some general recommendations the Danish municipalities must follow based on the ensuing effects stemming from this case.

## II. Background

The Danish DPA decision to ban the use of Google Chromebooks and Workspace in Helsingør Municipality is rooted in the fundamental principles and requirements of the GDPR, especially those related to lawful processing, transparency, accountability, data controllers' responsibilities and their duty to conduct a data protection impact assessment (DPIA).[5] This decision is also rooted in Articles 44 and 46(1) of the GDPR regarding the general principles for data transfers and transfers subject to appropriate safeguards.

Therefore, the Court of Justice of the European Union (CJEU) judgment in *Schrems II*[6] and its aftermath regulatory developments such as the European Data Protection Board (EDPB) Recommendations 01/2020[7] and the new Standard Contractual Clauses (SCCs)[8] for the transfer of personal data to third countries[9] were also considered. The GDPR restricts the transfer of personal data to third countries or international organizations outside of the European Economic Area (EEA), unless strict requirements are met.[10] Data importers and exporters must

---

[5] See Articles 5(1)(a), 5(2), 24, 28(1) and 35(1) of the GPDR.
[6] C-311/18 Data Protection Commissioner v. Facebook Ireland Limited, Maxilimian Schrems (Schrems II) [2020] ECLI: EU:C:2020:559.
[7] EDPB Recommendations on measures that supplement transfer tools to ensure compliance with the EU level of protection of personal data, adopted on 10 November 2020 [hereinafter EDPB Recommendations], available at: https://edpb.europa.eu/our-work-tools/our-documents/recommendations/recommendations-012020-measures-supplement-transfer_en (accessed 9 August 2022).
[8] See Standard Contractual Clauses for data transfer between the EU and non-EU countries, available at: https://ec.europa.eu/info/law/law-topic/data-protection/international-dimension-data-protection/standard-contractual-clauses-scc_en (accessed 9 August 2022).
[9] European Commission implementing decision of 4 June 2021 on standard contractual clauses for the transfer of personal data to third countries pursuant to Regulation (EU) 2016/679 of the European Parliament and of the Council, C(2021) 3972 final.
[10] See Chapter V of the GDPR, namely i) adequacy decisions (Art. 45 GDPR); ii) appropriate safeguards, such as standard contractual clauses, binding corporate rules, codes of conducts, certification mechanisms and ad-hoc contractual clauses (Art. 46 GDPR), and; iii) derogations, e.g., consent (Art. 49 GDPR).

guarantee that personal data will be protected under essentially equivalent European standards.[11]

The *Schrems II* case concluded that U.S. law does not ensure the adequate level of protection prescribed under Article 45 of the GDPR regarding the fundamental rights of EU data subjects guaranteed by Articles 7, 8 and 47 of the EU Charter of Fundamental Rights (CFR).[12] The CJEU deemed that Section 702 of the Foreign Intelligence Surveillance Act (FISA) and Executive Order (EO) 12333[13] does not grant the necessary limitations and safeguards concerning the interferences authorized by U.S. law.[14]

Consequently, the CJEU invalidated the EU-US Privacy Shield Framework[15] and found that SCCs are a valid tool to enable GDPR-compliant transfers of personal data if "supplementary measures" are in place to compensate for the potential lack of data protection arising from the third country law and practices. The effects of the *Schrems II* decision significantly raised the bar for organizations looking to comply with this complex legal landscape.[16] It also created a great deal of uncertainty amongst regulators, legal scholars and data protection professionals because these supplementary measures were not specified in the judgement. The Court neither defined nor provided examples of effective supplementary measures. Thus, such characterization undermined the original clarity of the GDPR with regards to the required standards for the security of processing as well as the available tools for cross-border transfers of personal data.[17]

---

[11] Unstaran, E., 'International Data Transfers' in Eduardo Unstaran (ed) *European Data Protection: Law and Practice*, 2nd edn., (IAPP Publication, 2019), 527.
[12] The significance of the fundamental right to respect for private life protected by Article 7 CFR, and the fundamental right to the protection of personal data guaranteed by Article 8 CFR have been discussed in various CJEU case law. See, e.g., C-553/07 Rijkeboer [2009] ECLI:EU:C:2009:293, para 47; Joined Cases C-293/12 and C-594/12 Digital Rights Ireland and Others [2014] ECLI:EU:C:2014:238, para 53; C-131/12 Google Spain and Google [2014] ECLI:EU:C:2014:317, paras 53, 66 and 74.
[13] While Section 702 FISA deals with all 'electronic communication service provider', Executive Order (EO) 12333 refers to electronic surveillance.
[14] Corrales Compagnucci, M., Minssen, T., Seitz, C., & Aboy, M. (2020). Lost on the High Seas without a Safe Harbor or a Shield? Navigating Cross-Border Data Transfers in the Pharmaceutical Sector After Schrems II Invalidation of the EU-US Privacy Shield. *European Pharmaceutical Law Review*, *4*(3): 156.
[15] Commission Implementing Decision (EU) 2016/1250 of 12 July 2016 pursuant to Directive 95/46/EC of the European Parliament and of the Council on the adequacy of the protection provided by the EU-US Privacy Shield (notified under document C(2016) 4176.
[16] Corrales Compagnucci, M., Aboy, M., & Minssen, T. (2021). Cross-Border Transfers of Personal Data after Schrems II: Supplementary Measures and New Standard Contractual Clauses (SCCs) . *Nordic Journal of European Law*, *4*(2), 37.

Despite the shortcomings of the *Schrems II* case, recent regulatory attempts have been more helpful in shedding some light and improving the overall safeguards associated with cross-border transfers of personal data. These include the EDPB Recommendations and the new version of the SCCs, which clarified the organizational and technical measures that data importers and exporters must implement. The EDPB Recommendations suggest a six-step approach and include a more detailed explanation of supplementary measures, highlighting the baseline core measures already enshrined in Art. 32 of the GDPR, namely, pseudonymization and encryption.[18]

The EDPB Recommendations include several use cases to demonstrate what it considers effective supplementary measures. The only effective supplementary measures are those: i) where the data transferred is fully encrypted, and the keys are kept solely under the control of the data exporter or an entity trusted by the exporter in the EEA; ii) the data transferred is effectively pseudonymized, and; iii) relying on split or multi-party computation.[19] The EDPB expressed that in situations where the third country importer has access to the encryption keys (namely, access to the unencrypted data in cleartext) and the public authorities may have the competence to access the transferred data lawfully, the "EDPB is incapable of envisioning an effective technical measure".[20]

Among the main innovations emerging from the EDPB Recommendations and new SCCs, are the obligation to assess the third country's law and practice. The underlying transfer must be assessed on a case-by-case basis to establish whether personal data will be adequately protected. Therefore, organizations should essentially follow a risk-based approach and perform a Transfer Impact Assessment (TIA).[21] Data exporters – in collaboration with the importer, where necessary – must assess if the law and practice of the third country where the data is being transferred may impinge on the effectiveness of the appropriate safeguards of the Article 46 GDPR mechanism used in the context of the specific transfer.[22] In performing such assessment, they should consider various aspects of the third country legal system[23] e.g., whether public authorities can lawfully access personal data of individuals. Suppose the TIA impact analysis revealed that the Article 46 GDPR transfer tool is not "effective". In that case, data exporters and importers must identify and implement additional safeguards to enhance data protection on a case-by-case basis.[24]

---

[17] For a critical appraisal of "supplementary measures" in the *Schrems II* decision, see Corrales Compagnucci, M., Fenwick, M., Aboy, M., & Minssen, T., (2022). Supplementary Measures and Appropriate Safeguards for International Transfers of Personal Data after Schrems II, available at SSRN: https://ssrn.com/abstract=4042000 or http://dx.doi.org/10.2139/ssrn.4042000 (accessed 13 August 2022).
[18] Ibid.
[19] Ibid.
[20] EDPB Recommendations. Paragraphs 80, 94-95 (Use Case 6), 96-97 (Use Case 7).
[21] Step 3 of the EDPB Recommendations.

## III. The Origins and Facts of the Danish DPA Decision

The use of Google Chromebooks and Workspace is widespread across Denmark. However, specifically the Danish DPA has had a pending case in Helsingør Municipality.[25] The facts of the case date back to 2010, where several schools in various municipalities in Denmark started using Google Apps and years later Google Chromebooks and Workspace (formerly G-Suite for Education) granting students access to programs such as Gmail and YouTube, without the parents' knowledge, consent and supervision. These accounts were necessary to install the G-Suite software package – a Google product specifically aimed at the education of children and young people.[26]

When students used these programs with their school accounts, personal data such as their names and school and class information were processed and displayed on the platforms. Parents were not aware of this, thus they did not have the chance to correct or anonymize this information, as only the Municipality had access to user management panels.[27] The

---

[22] Corrales Compagnucci, M., Aboy, M., & Minssen, T. (2021). Cross-Border Transfers of Personal Data after Schrems II: Supplementary Measures and New Standard Contractual Clauses (SCCs). *Nordic Journal of European Law*, *4*(2), 37 – 47.
[23] Generally, those elements mentioned in Article 45(2) of the GDPR 29, such as the rule of law situation and respect for human rights in that third country.
[24] However, Essentially equivalentCorrales Compagnucci, M., Aboy, M., & Minssen, T. (2021). Cross-Border Transfers of Personal Data after Schrems II: Supplementary Measures and New Standard Contractual Clauses (SCCs) . *Nordic Journal of European Law*, *4*(2), 37 – 47.
[25] For the Danish DPA statement (in Danish), see: https://www.datatilsynet.dk/afgoerelser/afgoerelser/2022/jul/datatilsynet-nedlaegger-behandlingsforbud-i-chromebook-sag- (accessed 13 August 2022).
[26] Nović, M. and Walle R.A., (Hosts). (2022, July 19th). Google Got Schooled (Audio Podcast, Episode 6). In *Grumpy GDPR.*, available at: https://share.transistor.fm/s/ab82490c (accessed 13 August 2022).
[27] It is also worth mentioning that Danish law does not allow municipalities to give away children's data, particularly for repurposing without a legal basis. Nović, M. and Walle R.A., (Hosts). (2022, August 3rd).

Municipality had previously neither performed a risk assessment for the processing in question nor implemented the technical and organizational measures necessary to ensure an adequate level of protection, considering the likelihood or severity of the risks related to the use of social media using the created school accounts.[28]

In December 2019, the parent of a student filed a complaint to the Danish DPA claiming that the child has – without his knowledge and consent – created a YouTube account,[29] whereby the child's personal data could be published on YouTube. Helsingør Municipality confirmed the complaint but stated that the risk was low, and it was unlikely that this would represent a risk to the rights and freedoms of the students. At the same time, the Municipality informed the Danish DPA that they do not use Google Workspace's additional services, arguing that the Municipality had no obligation to carry out a DPIA.[30]

## IV. Reasoning and Key Findings of the Danish DPA Decision

In September 2021, the Danish DPA issued a decision[31] ordering Helsingør Municipality to perform a risk assessment and impact analysis of the Municipality's processing activity, which reflects the flows of personal data that the processing entails in primary schools using Google Chromebooks and Workspace. It is also worth mentioning that the Municipality uses Google Cloud EMEA Limited as a data processor regarding its use of Google Chromebooks and Workspace. Therefore, the Danish DPA suspended any transfer of personal data to the U.S. that Helsingør Municipality has instructed Google Cloud EMEA Limited to carry out as a data

---

Back to School (Audio Podcast, Episode 8). In *Grumpy GDPR.,* available at: https://share.transistor.fm/s/d1192fab (accessed 13 August 2022).

[28] GDPR Hub, Country Report – Denmark, available at: https://gdprhub.eu/index.php?title=Datatilsynet_(Denmark)_-_2020-431-0061 (accessed 13 August 2022).

[29] In fact, one of the students received a negative comment on his YouTube account. This incident created concerns and anxiety for the child and parents. Nović, M. and Walle R.A., (Hosts). (2022, July 19th). Google Got Schooled (Audio Podcast, Episode 6). In *Grumpy GDPR.,* available at: https://share.transistor.fm/s/ab82490c (accessed 13 August 2022).

[30] GDPR Hub, Country Report – Denmark, available at: https://gdprhub.eu/index.php?title=Datatilsynet_(Denmark)_-_2020-431-0061 (accessed 13 August 2022).

[31] The order was announced in accordance with the corrective powers given to supervisory authorities in Art. 58(2)(d) of the GDPR. See 'Alvorlig kritik af Helsingør Kommune I Chromebook-sag' (10 September 2021), available at: https://www.datatilsynet.dk/afgoerelser/afgoerelser/2021/sep/afgoerelse-vedroerende-brud-paa-persondatasikkerheden (accessed 14 August 2022).

processor for the Municipality. The data transfer was suspended until the Municipality could demonstrate that the provisions enshrined in Chapter V of the GDPR have been observed.[32]

According to Helsingør Municipality, they have implemented the SCCs as a transfer basis and prepared a separate TIA following the Danish DPA and EDPB Recommendations requirements. The TIA included an assessment of whether legislation and practice in the U.S. affect the effectiveness of the SCCs. The Municipality has ensured via its settings that personal data is only stored in data centers within the EU/EEA countries.[33]

Notwithstanding the above setting, personal data may be transferred to Google LLC in the U.S. as part of remote access in connection to additional support services, where necessary. Based on the TIA's statistics and other arguments from the data importer, the risk of a public authority accessing personal data in the U.S. is 50 per cent. Thus, in Helsingør Municipality's interpretation of the assessment, the risk is more theoretical than practical. According to the Municipality, it is improbable that they will be the target of a request. The number of national security letter requests is so low that it is statistically without relevance.[34]

It also appears from the TIA that personal data transferred to Google LLC in the U.S. will be available to Google LLC in plain text. Google uses several layers of encryption and pseudonymization to protect customers' data at rest in Google products, utilizing one or more encryption mechanisms such as AES256, with a common cryptographic library which incorporates the FIPS 140-2 validated module, BoringCrypto, to implement encryption consistently across almost all Google Cloud products. Nonetheless, this encryption does not impede Google personnel from accessing Helsingør Municipality's personal data since Google has the key to decrypt it. This means that Google in the U.S. or other Google entities outside the EU/EEA or third countries can access the Municipality's personal data with the approval from the applicable Google entity established in the EU (Google Ireland). This implies that the Municipality has no complete control over the access of personal data in the EU data center.[35]

---

[32] GDPR Hub, Country Report – Denmark, available at:
https://gdprhub.eu/index.php?title=Datatilsynet_(Denmark)_-_2020-431-0061 (accessed 13 August 2022).
[33] Ibid.
[34] Ibid.
[35] Ibid. TIA's section 3.4

Regarding the legislation and practice in the U.S. potentially impinging on the effectiveness of the SCCs, it appears from the TIA[36] that Google LLC may qualify as an electronic commutations service provider under to Section 702 FISA for its U.S. customers. According to the Municipality's TIA, however, "…there is a high likelihood that the data accessible to the Google LLC, is per se excluded from access under Section 702 FISA because it is data that is not transmitted by it but to it to prove a support service. Thus, it is a communication targeted to a "U.S. person" for which the intelligence searches are prohibited…".[37] In addition, "Helsingør Kommune's personal data does not comprise personal data about "U.S. Persons", and U.S. authorities are thus barred from accessing data under Section 702 FISA for this reason as well".[38] Hence, in the Municipality's opinion, it is unlikely that personal data stored in the EU data centers will be subject to Section 702 FISA.[39]

Based on the Municipality's risk assessment results and accompanying documentation submitted by the Municipality, the Danish DPA found, however, that the risk to the rights and freedoms of the registered pupils was too high, and the processing did not comply with the GDPR requirements on several points, including the transfer of data to third countries (in particular, the U.S.) for technical support without the adequate level of security and protection.[40] In addition to several other criticisms towards the Municipality's data processing practices, the Danish DPA decided to ban such processing for violating the GDPR and suspended any related data transfers to the U.S. until it is brought in line with the GDPR.[41]

One of the main risks identified was processing personal data for other purposes. According to the Danish DPA reference to the Municipality's own assessment, even though Google is acting as a processor and the Municipality is ensured by virtue of the data processing agreement signed with Google, which clearly states that data will not be used for other purposes – including marketing – it cannot be ruled out the possibility that Google may break the contractual obligations and nevertheless use personal data for marketing or other unintended purposes for which the Municipality has not given instructions. The risk is described as follows:[42]

---

[36] Ibid. TIA's section 4.1.1.
[37] See Alan Charles Raul, 'Why Schrems II Might Not Be a Problem for EU-US Data Transfers', (21 Dec 2020), available at: https://www.lawfareblog.com/why-schrems-ii-might-not-be-problem-eu-us-data-transfers (accessed 8 August 2022); see also Alan Charles Raul, 'Schrems II Concerns Regarding U.S. National Security Surveillance Do Not Apply to Most Companies Transferring Personal Data to the U.S. Under Standard Contractual Clauses' (23 Dec 2020), available at: https://bit.ly/2V9veez (accessed 8 August 2022).
[38] GDPR Hub, Country Report – Denmark, available at:
https://gdprhub.eu/index.php?title=Datatilsynet_(Denmark)_-_2020-431-0061 (accessed 13 August 2022).
[39] Ibid.
[40] European Data Protection Board (EDPB), The Danish DPA imposes a ban on the use of Google Workspace in Elsinore municipality, available at: https://edpb.europa.eu/news/national-news/2022/danish-dpa-imposes-ban-use-google-workspace-elsinore-municipality_en (accessed 9 August 2022).
[41] Ibid.

> "There is a risk that Google or other third parties use personal data about teachers and students for marketing or other purposes for which Helsingør Municipality, as data controller, does not want personal data to be processed. In particular, contact information, IP address and digital traces (general information) are relevant in this context. It is noted that this is personal data related to students who enjoy special protection according to the data protection rules, which is why the access to and processing of personal data about the students constitutes an additional element in relation to the risk assessment."[43]

Regarding the issue of transferring data to third countries, the risk assessment described the risk as follows:

> "There is a risk that personal data about students and teachers (basically ordinary personal data, but it cannot be ruled out that sensitive personal data will also be included) will be transferred to unsafe third countries without a necessary transfer basis and without assurance that the third country in question ensures similar data protection rights as in other EU countries."[44]

Although the Danish DPA stated that the Municipality carried out meticulous work to map how personal data is used in the school, the Danish DPA upheld that the Municipality infringed Articles 5(2), cf. 5(1)(a), 24, cf. 28(1) 35(1) and 44, cf. 46(1) of the GDPR. The Danish DPA concluded that the Municipality as the controller has not successfully assessed some specific risks in relation to the data processing construction as to the processing activities the controller is allowed to do as a public authority.[45] The risk assessment did not cover all the risk scenarios that may arise as a result of the data processor design and the system choices. In addition, the Municipality did not conduct sufficient testing of the scope and operation of the selected hardware and software utilized. Finally, the Municipality failed to document how they controlled Google's access to personal data.[46]

---

[42] Datatilsynet (Denmark), GDPRhub, available at:
https://gdprhub.eu/index.php?title=Datatilsynet_(Denmark)_-_2020-431-0061 (accessed 9 August 2022).
[43] Ibid.
[44] Ibid.
[45] European Data Protection Board (EDPB), The Danish DPA imposes a ban on the use of Google Workspace in Elsinore municipality, available at: https://edpb.europa.eu/news/national-news/2022/danish-dpa-imposes-ban-use-google-workspace-elsinore-municipality_en (accessed 9 August 2022).

Regarding cross-border data transfers to the U.S., the Danish DPA found that the Municipality, in its role as data controller, instructed its processor (Google Ireland) to transfer personal data to a sub-processor (Google LLC) in the U.S. The transfer was based on the SCCs pursuant to Art. 46(2)(c) of the GDPR. However, as previously examined in the *Schrems II* judgment, the Court clarified that the use of SCCs does not always constitute "an adequate means of ensuring the effective protection of the personal data transferred to the third country in question in practice. That is the case, in particular, where the law of that third country allows its public authorities to interfere with the rights of data subjects in relation to that data."[47] Therefore, the Danish DPA stated that if there was a risk of transfer to the U.S., and such transfer being subject to problematic legislation, the risk to the rights and freedoms of pupils could be high. Accordingly, the Municipality was obliged to ensure that additional safeguards were implemented to comply with Chapter V of the GDPR. The Municipality argued that the data were encrypted. Nevertheless, in this case, the Danish DPA did not consider encryption to be effective since the Municipality itself recognized that Google LLC may have access to the information in plain text. On that account, the Danish DPA found that this violated Article 44 in conjunction with Article 46(2)(c) of the GDPR.[48]

Finally, in a new decision on August 18th, 2022,[49] the Danish DPA decided to maintain the ban to the use of Google Chromebooks and Workspace. The Danish DPA dealt with the material provided by the Municipality and reached the conclusion that the Municipality still failed to document that it has reduced the high risks for the children in Helsingør Municipality's schools.[50]

---

[46] Datatilsynet (Denmark), GDPRhub, available at: https://gdprhub.eu/index.php?title=Datatilsynet_(Denmark)_-_2020-431-0061 (accessed 9 August 2022).
[47] Ibid.
[48] Ibid.
[49] Datatilsynet, 'Datatilsynet Fastholder Forbud I Chromebook-sag' (in English: *The Danish Data Protection Authority Maintains the Ban in the Chromebook case*) (18 August 2022), available at: https://www.datatilsynet.dk/presse-og-nyheder/nyhedsarkiv/2022/aug/ny-afgoerelse-datatilsynet-fastholder-forbud-i-chromebook-sag (accessed 20 August 2022).
[50] Ibid.

In the opinion of Helsingør Municipality, all relevant risks have been identified and handled. However, according to the Danish DPA the Municipality has not assessed the relevant risks that appear in the contract with the supplier itself, and other publicly known risks issues in Google's technology they have chosen. In addition, the risks that the Municipality has identified have not been sufficiently mitigated. While Helsingør Municipality has assessed that Google only acts as a data processor, the Danish DPA believes that Google acts in several areas as an independent data controller that process personal data for its own purposes.[51]

## V. Concluding Remarks

This was a complex and sensitive case because it involved personal data of a vulnerable group. Some of the risks related to personal data and the use of social media by minors are, for instance, uploading embarrassing or provocative videos of themselves or others, being exposed to inappropriate or upsetting content, libel, sexual harassment, cyberbullying, etc.[52]

The Danish DPA emphasized the need to bring the illegal actions to an end as soon as possible. Helsingør Municipality's responsibility is to correct and delete information in conformity with the decision. The Municipality must contact the parents of the registered pupils in order to implement the corrections, anonymization or deletion of the registered personal data.[53]

The Danish DPA established that many of the conclusions in this decision would likely apply to other municipalities using the same Google tools with a similar processing structure – especially in relation to unauthorized disclosure and transfer to unsafe third countries. Given the centrality of data for many municipalities today, they cannot ignore the Danish DPA decision. While the changes that Danish Municipalities must implement can be burdensome and costly, they are expected to take relevant actions and develop strategies to manage these legal risks. Their obligation is to assess whether there are similar problems, and if so, in cooperation with the suppliers, to get control of the processing so that the equipment and tools can be used in accordance with the law.[54]

---

[51] Ibid.
[52] See, e.g., Stair, R., and Reynolds, G., '*Fundamentals of Information Systems*', 9th edn., (Cengage Learning, 2018), 455; Sandbrooks, J., and Brown, B., 'Social Media Used Among 9- to 11-Year Old Children and School Principals' Leadership Practices' in Marlynn Griffin and Cordelia Zinskie (eds), *Social Media: Influences in Education*, (Information Age Publishing Inc., 2021), 69; Tobin J., and Handsley, E., 'The Mass Media and Children: Diversity of Sources, Quality of Content, and Protection against Harm' in John Tobin (ed) *The UN Convention on the Rights of the Child: A Commentary* (Oxford University Press, 2019), 636.
[53] Ibid.

The are several ways by which they could mitigate the practical challenges that such transfers created post-Schrems. First and foremost, Danish municipalities should carry out a TIA demonstrating that the transfer in question is not subject to FISA 702 or other problematic legislation.[55] Moreover, while this could increase costs, it would be beneficial if organizations could invest in risk assessment software tools, following the privacy by design and by default approach.[56]

In this context, it is worth noting that contractual and organizational measures are not enough. They will not render ineffective access to personal data by U.S. law enforcement authorities. The most effective supplementary measures mainly, when using cloud computing services – are technical measures such as full encryption,[57] pseudonymization and so-called split processing.[58]

---

[54] Datatilsynet, 'Datatilsynet Fastholder Forbud I Chromebook-sag' (in English: *The Danish Data Protection Authority Maintains the Ban in the Chromebook case*) (18 August 2022), available at: https://www.datatilsynet.dk/presse-og-nyheder/nyhedsarkiv/2022/aug/ny-afgoerelse-datatilsynet-fastholder-forbud-i-chromebook-sag (accessed 20 August 2022).

[55] See, e.g., Walle R.A., 'Ultimate Resources for SCCs and TIAs – Schrems II' (), available at: https://www.noties.consulting/ultimate-resources-gdpr-sccs-tias-schrems-ii/ (accessed 15 August 2022); see also, Rosenthal, D., 'Frequently Asked Questions (FAQ) on the Risk of Foreign Lawful Access and the Statistical "Rosenthal" Method for Assessing It' (1 August 2022), available at: https://www.rosenthal.ch/downloads/Rosenthal-LA-method-FAQ.pdf (accessed 15 August 2022).

[56] See, e.g., Djemame, K., et al., (2013). Legal Issues in Clouds: Towards a Risk Inventory. *Philosophical Transactions of the Royal Society A: Mathematical, Physical and Engineering Sciences*, *371*(1983), 1-17. https://doi.org/10.1098/rsta.2012.0075

[57] See, e.g., Barnitzke, B., et al., (2011). Legal restraints and security requirements on personal data and their technical implementation in clouds. In *Workshop for E-contracting for Clouds, eChallenges*, 51-55; Kiran, M., et al. (2012). Managing security threats in Clouds. *Digital Research*; T. Kirkham et al. (2012). Assuring Data Privacy in Cloud Transformations, *2012 IEEE 11th International Conference on Trust, Security and Privacy in Computing and Communications*, 2012, 1063-1069, doi: 10.1109/TrustCom.2012.97; Badia, R.M. *et al.* 'Demonstration of the OPTIMIS Toolkit for Cloud Service Provisioning' in: Abramowicz, W., et al (eds) *Towards a Service-Based Internet. ServiceWave 2011*. Lecture Notes in Computer Science, vol 6994 (Springer, 2011).

[58] Danish DPA (*Datatilsynet*), 'Guidance on the Use of Clouds' (March 2022), 22, available at: https://www.datatilsynet.dk/Media/637824108733754794/Guidance%20on%20the%20use%20of%20cloud.pdf (accessed 9 August 2022).

Hence, one approach is to employ technical solutions such as multiparty homomorphic encryption.[59] This approach combines secure multiparty computation and allows to query and analyze encrypted data without the necessity to decrypt the data. Thus, allowing municipalities to overcome their respective limitations and provide scalable and effective data protection measures.[60]

Another alternative is exploring emerging technology-driven solutions allowing organizations to store data in the end-user's personal cloud. This approach follows the 'user-held' data model that grants individuals greater control and provides more value over their data. One advantage of this approach is that it significantly reduces the need for unnecessary open-ended cross-border data transfers, as users could choose their geographic location and keep their data stored locally within the EEA in a more decentralized manner.[61]

**Acknowledgement:** The research for this paper was supported by a Novo Nordisk Foundation grant for a scientifically independent Collaborative Research Program in Biomedical Innovation Law (grant agreement number NNF17SA0027784).

---

[59] For a full account of homomorphic encryption, see: Corrales Compagnucci, M., et al., (2019). Homomorphic Encryption: The 'Holy Grail' for Big Data Analytics & Legal Compliance in the Pharmaceutical and Healthcare Sector? *European Pharmaceutical Law Review*, *3* (4), 144-155.

[60] Corrales Compagnucci, M., Fenwick, M., Aboy, M., & Minssen, T., (2022). Supplementary Measures and Appropriate Safeguards for International Transfers of Personal Data after Schrems II, available at SSRN: https://ssrn.com/abstract=4042000 or http://dx.doi.org/10.2139/ssrn.4042000 (accessed 13 August 2022).

[61] Jurcys, P., Corrales Compagnucci, M., & Fenwick, M. (2022). The Future of International Data Transfers: Managing New Legal Risk with a 'User-Held' Data Model. *Computer Law & Security Review*, *46*, [105691]. https://doi.org/10.1016/j.clsr.2022.105691